\documentclass[aps,prx,twocolumn,amsmath,amssymb,floatfix,longbibliography]{revtex4-1}

\usepackage[utf8]{inputenc}

\usepackage{graphicx}
\usepackage{dcolumn}
\usepackage{bm}
\usepackage{dsfont}
\usepackage{physics}
\usepackage{amsmath,amsthm,amsfonts,amssymb,amscd}
\usepackage[mathlines]{lineno}
\usepackage[usenames, dvipsnames]{color}
\usepackage{mathtools}
\usepackage{textcomp}
\usepackage{upgreek}
\usepackage{hyperref}
\hypersetup{
    colorlinks=true,
    linkcolor=blue,
    filecolor=blue,      
    urlcolor=blue,
    citecolor = blue
}

\newcommand{\um}{$\upmu\mathrm{m}\;$}

\newcommand{\umdot}{$\upmu\mathrm{m}$}

\newcommand*{\figref}[2][]{%
  Fig.~\hyperref[{fig:#2}]{%
    \ref*{fig:#2}%
    \ifx\\(#1)\\%
    \else
      (#1)%
    \fi
  }%
}

\usepackage{bbold}
\usepackage{titlesec}
\titlespacing*{\section} {0pt}{3.5ex plus 2ex minus .2ex}{2.3ex plus .2ex}

\usepackage{ulem}

\begin{document}
\preprint{}
\title{Resonant Excitation and Purcell Enhancement of Coherent Nitrogen-Vacancy Centers Coupled to a Fabry-P\'{e}rot Micro-Cavity}

\author{M. Ruf$^{1,2}$}\thanks{These authors contributed equally to this work.}  
\author{M. J. Weaver$^{1,2}$}\thanks{These authors contributed equally to this work.}
\author{S. B. van Dam$^{1,2}$}
\author{R. Hanson$^{1,2}$}
\thanks{Correspondence and requests for materials should be addressed to m.t.ruf@tudelft.nl or r.hanson@tudelft.nl}

\date{\today}

\affiliation{$^{1}$QuTech, Delft University of Technology, PO Box 5046, 2600 GA Delft, The Netherlands}
\affiliation{$^{2}$Kavli Institute of Nanoscience Delft, Delft University of Technology, PO Box 5046, 2600 GA Delft, The Netherlands}

\date{\today}

\begin{abstract} 

 The nitrogen-vacancy (NV) center in diamond has been established as a prime building block for quantum networks. However, scaling beyond a few network nodes is currently limited by low spin-photon entanglement rates, resulting from the NV center's low probability of coherent photon emission and collection. Integration into a cavity can boost both values via the Purcell effect, but poor optical coherence of near-surface NV centers has so far prevented their resonant optical control, as would be required for entanglement generation. Here, we overcome this challenge, and demonstrate resonant addressing of individual, fiber-cavity-coupled NV centers, and collection of their Purcell-enhanced coherent photon emission. Utilizing off-resonant and resonant addressing protocols, we extract Purcell factors of up to 4, consistent with a detailed theoretical model. This model predicts that the probability of coherent photon detection per optical excitation can be increased to 10\% for realistic parameters - an improvement over state-of-the art solid immersion lens collection systems by two orders of magnitude. The resonant operation of an improved optical interface for single coherent quantum emitters in a closed-cycle cryogenic system at $T \sim$ 4 K is an important result towards extensive quantum networks with long coherence.
 
\end{abstract}

\maketitle

\section{INTRODUCTION}

Future large-scale quantum networks sharing entanglement between their nodes may enable a suite of new applications, such as secure communication, distributed quantum computation, and quantum enhanced sensing~\cite{Kimble2008,Wehner2018,Ekert2014, Nickerson2014,Gottesman2012}. These networks require nodes with both access to long-lived memory qubit registers that can be operated with high fidelity, and bright spin selective optical transitions with good coherence~\cite{Wehner2018}. Promising node candidates include group IV defects in diamond~\cite{Nguyen2019,Bhaskar2020,Trusheim2020}, defects in SiC~\cite{Christle2017,Nagy2019,Bourassa2020}, rare earth ions in solid state hosts~\cite{Raha2020, Kindem2020,Merkel2020}, quantum dots~\cite{Delteil2016,Stockill2017}, neutral atoms~\cite{Ritter2012,Hofmann2012}, and trapped ions~\cite{Hucul2015,Stephenson2020}.

The nitrogen-vacancy (NV) center in diamond combines optical transitions suitable for remote entanglement generation under moderate cryogenic conditions with outstanding electron spin coherence ($T_2$ $>$ 1 s) and extensive control capabilities over local $^{13}$C memory atoms \cite{Bernien2013,Bradley2019}; these features have enabled pioneering quantum network experiments~\cite{Kalb2017,Humphreys2018} and fundamental tests of physics~\cite{Hensen2015}. However, entanglement generation rates are limited by the relatively low photon emission into the zero-phonon line (ZPL), as well as low collection efficiency from diamond, hindering scaling beyond a few nodes. Both values can be significantly increased by embedding the NV center inside an optical cavity, making use of the Purcell effect. Cavity-coupling of NV centers at low temperature has been demonstrated for different cavity implementations, including photonic crystals~\cite{Faraon2012,Hausmann2013,Li2015,Riedrich-Moller2015,Schroder2017a,Jung2019}, microdisk resonators~\cite{Faraon2011}, and open micro-cavities~\cite{Riedel2017a,Johnson2015}. 

Entanglement generation between nodes can be achieved through resonant excitation and photon detection from individual, coherent emitters~\cite{Bernien2013}. However, poor optical NV center coherence ($\sim$ GHz linewidths), resulting from surface noise effects and/or implantation-induced damage, has prevented the required resonant optical addressing of Purcell-enhanced NV centers in optical cavities to date~\cite{Faraon2012,Riedrich-Moller2015,Riedel2017a,Jung2019,VanDam2019,Kasperczyk2020}. 

In this work, we capitalize on recent breakthroughs in diamond membrane fabrication~\cite{Ruf2019} and diamond-based open micro-cavities~\cite{Janitz2015,Bogdanovic2017,BogdanovicFab2017,Riedel2017a,Haussler2019,Jensen2020,Salz2020} and demonstrate Purcell enhancement of coherent, resonantly excited NV centers coupled to a fiber-based cavity. We outline the experimental system in Sec.~\ref{sec:exp_setup_main}, and begin experiments by verifying Purcell enhancement under off-resonant excitation in Sec.~\ref{section:Off-resonant}. Next, we develop a resonant excitation protocol for measuring optical coherence (Sec.~\ref{sec:line_scans}) and Purcell enhancement (Sec.~\ref{sec:Resonant_purcell}). Finally, in Sec.~\ref{sec:ZPL_collection}, we detect coherent, Purcell enhanced photon emission from a single quantum emitter, and model current photon loss sources, paving the way for future experimental designs.

\begin{figure*}
    \centering
    \includegraphics[width=.99\textwidth]{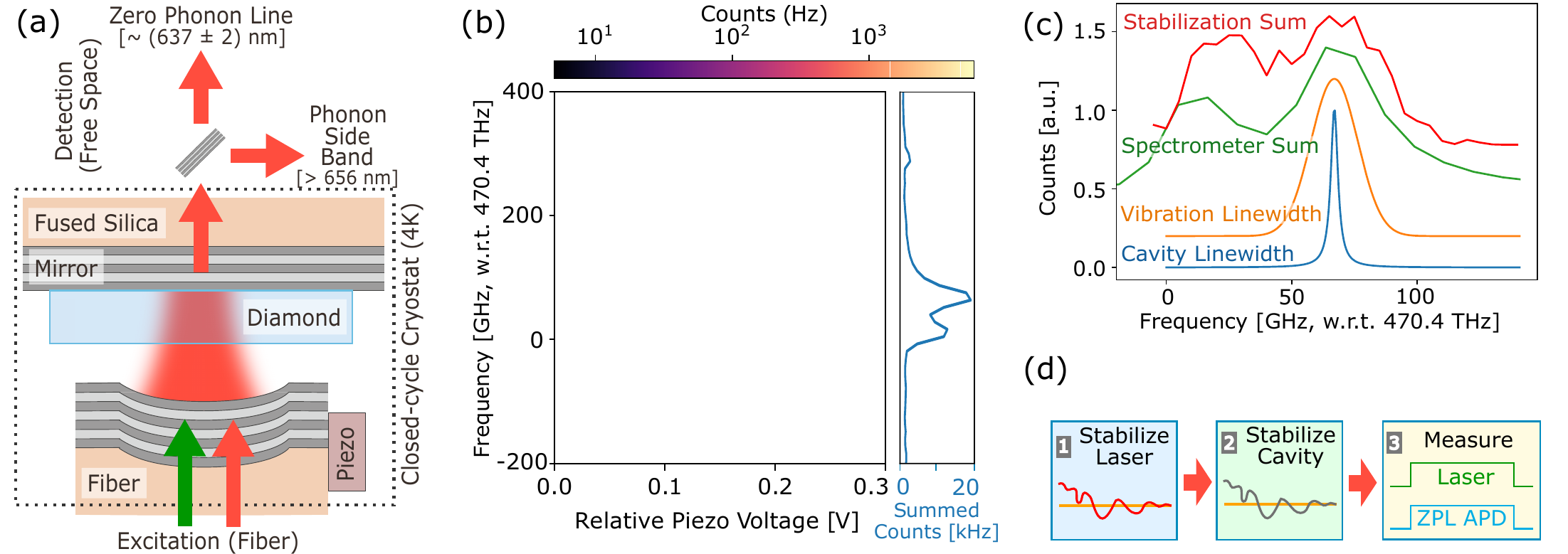}
    \caption{Experimental setup, NV-cavity characterization measurements, and measurement sequence. (a) Overview of the experimental setup. A tunable fiber-based micro-cavity with embedded diamond membrane is formed inside a closed-cycle cryostat operated at a temperature of 4 K. (b) Fluorescence counts for different applied piezo voltages (and thus cavity lengths) under off-resonant excitation, detected with a spectrometer in a $\pm$ 2 nm window around the NV center zero-phonon line (ZPL) of $\sim$ 637 nm (integration time 5 s per point). Right panel shows data on the left, summed over all piezo voltages. (c) From bottom to top, measurements of intrinsic cavity linewidth, inferred vibrations-broadened cavity linewidth, spectrometer peaks (same data as right panel of (b)), and stabilization curve, using the continuous off-resonant excitation and ZPL detection measurement sequence displayed in (d). (d) General measurement sequence used throughout this paper. A $\sim$ 637 nm laser is frequency stabilized to a wavemeter (1), and serves as a reference for the cavity length (2). Measurement blocks that are repeated multiple times are interleaved between stabilization rounds (3). }
    \label{fig:Overview}
\end{figure*}

\section{EXPERIMENTAL SETUP}\label{sec:exp_setup_main}

An overview of the experimental setup used in this work can be seen in \figref[a]{Overview}. At the heart of the experiment is an open, fiber-based Fabry-P\'{e}rot micro-cavity (design finesse 6200), formed from a flat, super-polished mirror, and a laser ablated fiber mirror~\cite{Hunger2010}; this finesse value is chosen as it maximises the outcoupled fraction of photons emitted into the zero-phonon line (ZPL) for the vibrations present in the cavity~\cite{VanDam2018} ($\sim$ [0.1 - 0.2] nm root mean squared amplitude, for a detailed discussion, please see the supplementary material~\cite{supplement}). The fiber sits on top of a piezo positioning stage, which enables in situ tuning of the cavity position and length under operation in a closed-cycle cryostat ($T$ $\sim$ 4 K for all measurements in this paper). Excitation light is delivered to the cavity via the fiber mirror, while all detection in this work takes place in free space through the flat mirror of the cavity. Unbalanced mirror coatings set the design finesse almost entirely by transmission through this flat cavity mirror; for a full overview of the experimental setup, please see~\cite{supplement}. An electron irradiated and annealed diamond membrane is bonded to the flat cavity mirror~\cite{BogdanovicFab2017}, and etched down to a final thickness of $\sim$ 5.8 \um in the cavity region, following the process flow developed in Ref.~\cite{Ruf2019}. Importantly, this recipe has been shown to preserve the optical coherence of NV centers needed for entanglement generation, even for few-\um thin diamond samples.

We start characterizing the coupled system of the membrane and fiber-cavity by recording cavity spectra under illumination with a broadband light source for different cavity lengths. From fits to a transfer matrix model~\cite{Janitz2015}, we infer a membrane thickness of $\sim$ 5.8 \umdot, and an air gap of $\sim$ 7 \um (see App.~\ref{sec:cav_char}). This air gap could not be reduced further, likely due to a piece of dust on, or an angled mounting of, the fiber. This limits the cavity finesse to $\sim$ 2000 due to operation in the clipping loss regime of the fiber mirror~\cite{Hunger2010,VanDam2018}. The cavity parameters (NV center excited state decay rate, NV-cavity coupling rate, and cavity decay rate \{$\gamma, g, \kappa$\} $\sim$ $2\pi$ $\times$ \{13 MHz, 180 MHz, 3.5 GHz\}) place the system in the weak coupling regime of cavity quantum electrodynamics, in which collection of photons from an NV center is maximized (App.~\ref{sec:cav_char}). 

We find NV centers by scanning a fundamental cavity mode over the ZPL transition frequencies ($\sim$ 470.4 THz), while constantly illuminating with off-resonant ($\sim$ 515 nm) laser light. At this laser wavelength, the optical excitation keeps the NV center predominantly in the negative charge and $m_s$ = 0 spin state. \figref[b]{Overview} shows the fluorescence counts in a window of $\pm$ 2 nm around the expected ZPL frequency on a spectrometer, for different piezo voltages (and thus cavity lengths). Background light in the cavity serves as an internal light source~\cite{Riedel2020}, revealing the expected decrease in cavity frequency with increasing voltage. As the cavity mode is tuned through the NV center ZPL transition frequencies, additional NV center fluorescence is collected. Summing the spectrometer fluorescence counts per frequency over all cavity lengths reveals two peaks (right panel of \figref[b]{Overview}), each significantly wider than the intrinsic cavity linewidth.

To investigate the origin of the observed width of the NV center emission peaks, we compare the intrinsic cavity linewidth (blue trace in \figref[c]{Overview}) - measured on a timescale much faster than the vibrations in the cavity~\cite{supplement} - to the ``vibration linewidth", calculated by convolving the measured vibration value with the intrinsic cavity linewidth (orange trace in \figref[c]{Overview}). The resulting curve, which is an estimate for the linewidth averaged over during the full spectrometer run, explains most of the width of the peaks in the summed spectrometer data (green trace in \figref[c]{Overview});  the origin of the additional widening of the peaks is investigated below.  

For more refined cavity control, we design a measurement protocol which we use throughout the remainder of this paper (see \figref[d]{Overview}). The sequence starts by stabilizing a $\sim$ 637 nm laser to a given setpoint using a wavemeter. We then stabilize the cavity to this laser frequency, after which we start a measurement sequence. By repeating this procedure, we can take out slow drifts between measurements. We test this measurement protocol by stabilizing the cavity to different laser setpoints and taking fluorescence data in the ZPL under off-resonant excitation (red trace in \figref[c]{Overview}). The resulting data is consistent with the lineshape found via the spectrometer measurement. 

\section{OFF-RESONANT EXCITATION} \label{section:Off-resonant}

\begin{figure}
    \centering
    \includegraphics[width=0.45\textwidth]{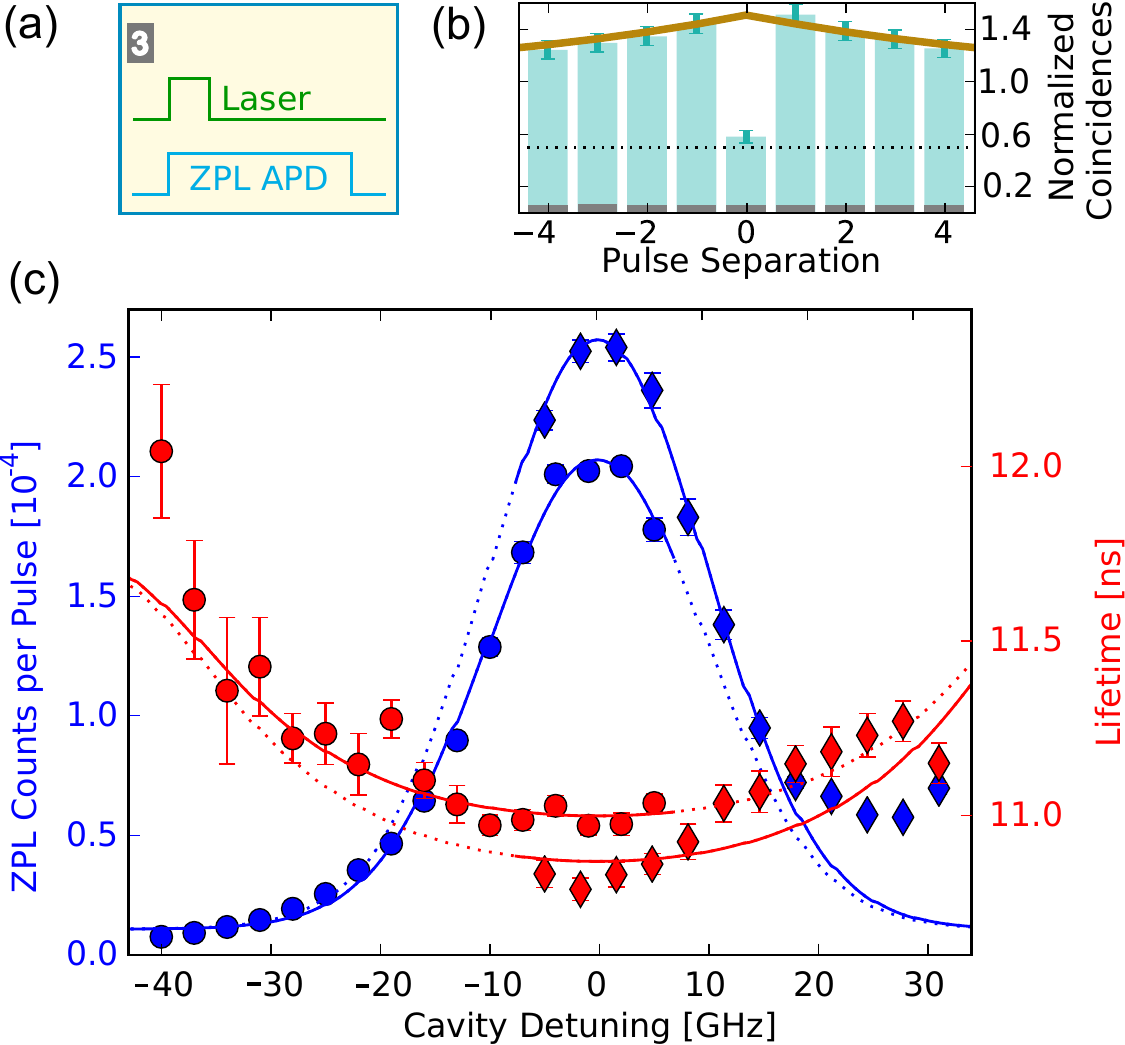}
    \caption{Purcell enhancement of NV centers under off-resonant excitation. (a) Measurement block for pulsed off-resonant excitation and detection in the ZPL. (b) Normalized pulsed autocorrelation measurement of NV center fluorescence from the cavity on two ZPL detectors. Dark boxes on the bottom are background fluorescence between the pulses. We fit the data to a simple bunching model based on probabilistic state initialization into a bright state (solid line)~\cite{supplement}. (c) Detuning sweep of the cavity with respect to the NV centers with measured ZPL fluorescence counts (blue) and lifetime (red). The data is taken on different days (circles and diamonds). We perform a joint fit of our model to both curves (solid lines)~\cite{supplement}. (b) and (c) were measured at different locations on the sample, which are also different from Fig.~\ref{fig:Overview}.}
    \label{fig:Offresonant}
\end{figure}

Entanglement generation rates in quantum networks scale with the collection of coherent photon emission. A cavity acts as a spectral filter, only allowing resonant emission to exit, and opens up an additional decay channel for the excited state of any coupled NV centers, which in turn decreases their lifetime. Thus, when the cavity is resonant with an NV center optical transition, more ZPL light should be emitted with a reduced lifetime, as has been observed in different cryogenic systems~\cite{Riedel2017a,Faraon2012,Hausmann2013,Li2015,Riedrich-Moller2015,Schroder2017a,Jung2019,Faraon2011}. Importantly, this additional decay is also funneled into the cavity mode which can be readily collected, as opposed to free space systems, in which the ZPL light is emitted in all directions. From the reduced lifetime, one can (following the terminology of Ref.~\cite{Riedel2017a}) extract the Purcell enhancement induced by the cavity, $F_P^{ZPL}$, as
\begin{equation}
    F_P^{ZPL} = \frac{1}{\beta_0}\left(\frac{\tau_0}{\tau'}-1\right) + 1,
\label{eq:FZPL}
\end{equation}
where $\beta_0$ is the Debye-Waller factor, recently estimated to be 2.55\%~\cite{Riedel2017a}, $\tau_0$ is the lifetime without the influence of the cavity, and $\tau'$ is the reduced NV center lifetime (see App.~\ref{section:theory} for a derivation). In this paper, we choose the definition of the Purcell factor as the increased emission into the zero-phonon line (ZPL) only - rather than the increase in total emission - because it better reflects the coherent ZPL light which can be used for entanglement. A subset of the ZPL light, $(F_P^{ZPL}-1)\beta_0/(F_P^{ZPL}\beta_0+1-\beta_0)\approx(F_P^{ZPL}-1)\beta_0$, is collectable via the cavity mode; this definition means that $F_P^{ZPL}$ = 1 without Purcell enhancement from the cavity. 

To measure lifetime and fluorescence from NV centers in the cavity under off-resonant excitation, we replace the measurement block in \figref[d]{Overview} with pulsed green excitation, pictured in \figref[a]{Offresonant}. The cavity is again stabilized with a red laser and the excitation is provided by a $\sim$ 532 nm picosecond pulsed laser input through the fiber. We collect the resulting fluorescence in the ZPL path on an avalanche photodiode (APD), and extract counts per pulse and lifetime as we sweep the detuning of the cavity. For further experimental details please see the supplemental material~\cite{supplement}.

\begin{figure*} 
    \centering
    \includegraphics[width=1.\textwidth]{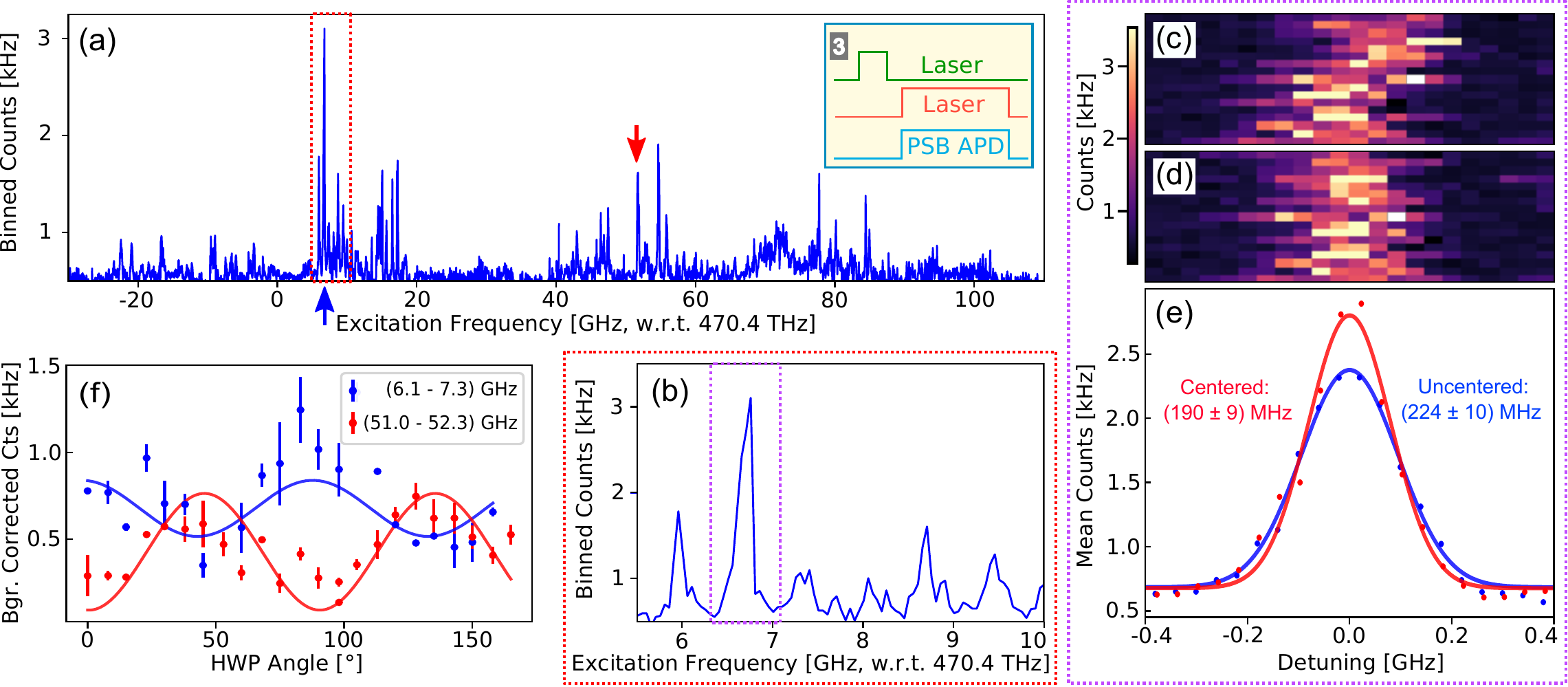}
    \caption{Photoluminescence excitation (PLE) scans. (a) Broad PLE scan (cavity resonant with excitation laser, binned into 50 MHz bins), revealing multiple bright lines per cavity position. Blue and red arrows indicate the center position of polarization angle scans displayed in (f). (Inset, a) Measurement block used during PLE scans. (b) Zoom-in of the red dashed frequency region in (a). (c) Series of 17 consecutive PLE scans (measurement time $\sim$ 15 min per trace) for the purple shaded frequency region in (b). (d) Same data as in (c), but center position of Gaussian fit to each individual trace used to correct for slow drift during the measurement. (e) Averaged and Gaussian fitted data of (c, blue) and (d, red), showing spectral diffusion limited linewidths of (224 $\pm$ 10) MHz and (190 $\pm$ 9) MHz, respectively. (f) Background corrected mean fluorescence counts during PLE scans for a frequency region of (6.1 - 7.3) GHz (blue) and (51. - 52.3) GHz (red, each w.r.t. 470.4 THz) as a function of excitation laser half wave plate (HWP) angle, showing that individual transitions are polarized. Solid lines are sinusoidal fits. }
    \label{fig:PLE}
\end{figure*}

\figref[c]{Offresonant} shows one such detuning sweep, in which the fluorescence counts and lifetime vary with cavity detuning. The highest fluorescence counts and lowest NV center lifetime coincide, demonstrating Purcell-enhanced NV center emission induced by coupling to the cavity. The widths of the fluorescence peak and of the lifetime reduction curve are several times broader than the cavity linewidth. To understand the quantitative behavior of the detuning sweep, we introduce a model which includes vibrations of the cavity and a spectral distribution of NV center transition frequencies (see App.~\ref{section:theory}). By fitting this model to the data, we can extract the emission into the ZPL in the cavity, $(F_P^{ZPL}-1)\beta_0$ = (7.9 $\pm$ 2.2)\%, and the off-resonant lifetime, $\tau_0$ = (11.8 $\pm$ 0.2) ns, which is consistent with the lifetime of NV centers reported for bulk diamond~\cite{Robledo2011,Kalb2018}. This data was taken on two different days, leading to two different curves in \figref[c]{Offresonant} (circles and diamonds) that can be explained by drifts, which we account for in our modeling~\cite{supplement}.

We investigate whether the emission is produced by a single emitter using an autocorrelation ($g^{(2)}$) measurement. At most locations, there is little or no drop in coincidence counts at zero pulse separation. Therefore, we conclude that we are addressing several emitters within the cavity mode volume, likely because the high density of NV centers in our sample lowers the chance of single center addressing with off-resonant excitation. \figref[b]{Offresonant} displays the most significant drop in coincidences observed, which falls to (0.58 $\pm$ 0.05) at zero time delay ([0.54 $\pm$ 0.05] with background correction). We observe significant bunching behavior for small pulse separations, which we attribute to probabilistic state initialization into a bright state~\cite{supplement}. The Purcell enhancement of the ZPL by a factor of 3.9 $\pm$ 0.9 (assuming the same $\beta_0$ as in Ref.~\cite{Riedel2017a}), is a lower bound for the largest enhancement of single centers at this spot, because there are multiple NV centers within the cavity mode volume.

\section{CONTINUOUS RESONANT EXCITATION}\label{sec:line_scans}

Entanglement generation in quantum networking protocols requires coherent addressing of individual zero-phonon line (ZPL) transitions with linewidths close to their lifetime-limited value~\cite{Bernien2013}. To determine the suitability of our device for such tasks, we now move on to photoluminescence excitation (PLE) scans. 

The measurement protocol for these scans is displayed in \figref[a, inset]{PLE}: short green pulses used for initialization of the NV center in the negative charge and m$_s$= 0 spin state are interleaved with red measurement pulses. Fluorescence counts collected in the phonon-sideband (PSB) during the red pulses are then correlated with simultaneously recorded wavemeter readings. This scheme thus allows predominant detection of the two m$_s$ = 0 spin-conserving optical transitions per NV center; these transitions connect the ground state to the two optically excited states typically labeled as E$_x$ and E$_y$. For further details about this measurement sequence, please see the supplemental material~\cite{supplement}.

A resulting PLE scan can be seen in \figref[a]{PLE}. We observe a multitude of narrow lines per cavity spot, confirming our interpretation that there are multiple NV centers present per cavity mode volume. \figref[b]{PLE} shows a zoom-in into the red dashed region of \figref[a]{PLE}; note that each binned point is comprised of many underlying individual datapoints. Importantly, the individual transition peaks can be spectrally distinguished, which is a pre-requisite for single resonant NV center addressing, as probed below.

To test the spectral stability of the transition peaks, we repeatedly scan the excitation laser over a NV center ZPL transition. \figref[c]{PLE} shows a series of 17 consecutive scans over the frequency region dashed in purple in \figref[b]{PLE}. To correct for slow drifts in this measurement (likely due to temperature fluctuations in the cryostat), we fit a Gaussian lineshape to each individual PLE trace, and shift the lines to a common center frequency, displayed in \figref[d]{PLE}. \figref[e]{PLE} shows the averaged data of the original and centered scans in blue and red, respectively. From Gaussian fits to this data (solid lines), we extract full width at half maximum linewidths of (224 $\pm$ 10) MHz and (190 $\pm$ 9) MHz for the original and centered case, respectively. We probed the linewidths of 14 NV center transitions during the course of this study, for a total of 4 different cavity positions. The supplemental material~\cite{supplement} shows the individual linewidths of each NV center transition, which average to (204 $\pm$ 59) MHz and (168 $\pm$ 49) MHz, for the un-centered and centered case, respectively. Thus, the linewidths in our sample are comparable to the ones found in bulk diamond samples for green repumping~\cite{Bernien2012}. Importantly, NV centers with similar optical coherence have enabled previous entanglement generation experiments~\cite{Bernien2013}.

We next investigate the polarization behaviour of NV center transitions in the cavity by scanning the polarization angle of the excitation laser with a half wave plate (HWP), and observe a dependence of the background corrected fluorescence counts on this HWP angle, see \figref[f]{PLE}; polarization can be used to suppress resonant excitation pulses in a cross-polarization detection scheme~\cite{Bernien2013}, as demonstrated below. Interestingly, the NV centers in the two studied frequency clusters around 470.407 THz and 470.452 THz show a different polarization behaviour. A possible origin of this effect is a strain-induced splitting in transition frequencies for E$_x$ and E$_y$ polarized lines, as observed in Ref.~\cite{Ruf2019}; future investigation is required to conclusively determine the origin~\cite{supplement}. 

\section{PULSED RESONANT EXCITATION}\label{sec:Resonant_purcell}
\label{sec:resonant exc}

\begin{figure}
    \centering
    \includegraphics[width=0.47\textwidth]{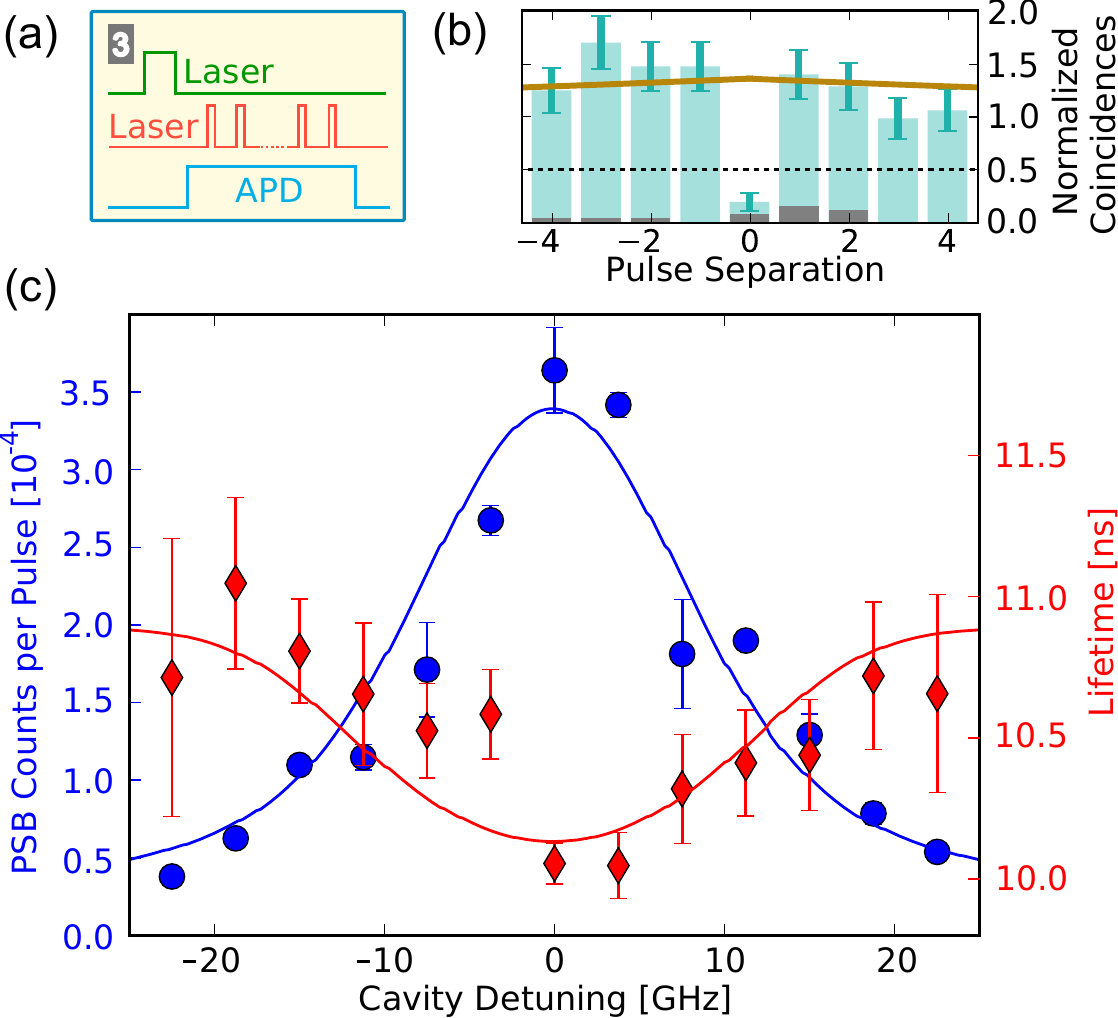}
    \caption{Purcell enhancement of NV centers probed with resonant excitation. (a) Measurement block used for pulsed resonant excitation and detection in the PSB / ZPL . (b) Normalized pulsed autocorrelation measurement of NV center fluorescence from the cavity on two PSB detectors. Dark boxes on the bottom are background fluorescence measured with the excitation detuned by 5 GHz. The fit accounts for bunching effects due to probabilistic state initialization into a bright state and the finite pulse train we apply (solid line)~\cite{supplement}. (c) Detuning sweep of the cavity with respect to the NV center with measured PSB fluorescence (blue) and lifetime (red). We perform a joint fit of both curves to a model (solid lines) with four free parameters~\cite{supplement}. (b) and (c) were measured on the same NV center as \figref[c-e]{PLE}, which is a different emitter than the ones studied in \figref[a-b]{Offresonant}.}
    \label{fig:ResonantPurcell}
\end{figure}

Now that we can resolve individual NV centers spectrally, we characterize their Purcell enhancement with a detuning sweep similar to that of Sec.~\ref{section:Off-resonant}. The measurement block replaces off-resonant pulses with single frequency resonant pulses (\figref[a]{ResonantPurcell}). The sequence consists of a green repump pulse to initialize the NV center predominantly in the negative charge and m$_s$ = 0 spin state, followed by a series of short ($\sim$ 2 ns) red pulses at the frequency of the NV center transition. We record the fluorescence counts in the phonon-sideband (PSB) path after each pulse, and extract the NV center lifetime. For further details about the measurement sequence, please see the supplementary material~\cite{supplement}.

First, we perform a pulsed autocorrelation measurement with two detectors in the PSB path to confirm that the individual peaks measured during photoluminescence excitation scans are indeed from single NV centers. \figref[b]{ResonantPurcell} displays the normalized g(2) value of (0.19 $\pm$ 0.09) ([0.16 $\pm$ 0.07] with background correction) for zero pulse separation. Unlike the off-resonant case, the value at zero pulse difference clearly falls below 0.5 in all three NV centers we tested, indicating that we are observing single quantum emitters.

For the same NV center measured in \figref[c-e]{PLE}, we sweep the detuning of the cavity from the NV center transition while keeping the excitation pulses resonant with the latter to measure the Purcell enhancement. The collection efficiency of PSB emission is independent of the cavity length. However, the probability of exciting the NV center depends on the overlap between the excitation laser frequency and the cavity resonance, so the PSB intensity should vary with detuning for fixed excitation power (see App.~\ref{section:theory}). In the measurement (\figref[c]{ResonantPurcell}), the fluorescence counts increase and the lifetime decreases when the cavity is on resonance with the NV center, demonstrating that we observe Purcell-enhanced NV center emission. We fit our model to both curves and extract the fraction of ZPL emission into the collectable cavity mode, $(F_P^{ZPL}-1)\beta_0$ = (7.0 $\pm$ 3.4)\% (see definition in Sec.~\ref{section:Off-resonant}), the off-resonant lifetime, $\tau_0$ = (10.9 $\pm$ 0.2) ns, and the root mean squared cavity vibrations, $\sigma_{vib}$ = (0.18 $\pm$ 0.02) nm~\cite{supplement}. The Purcell enhancement of this NV center is consistent with the enhancement we found for NV centers under off-resonant excitation.

\section{ZPL COLLECTION AND FUTURE IMPROVEMENTS}\label{sec:ZPL_collection}
\label{sec:zplcollectionfurtherimprov}

So far, we have only studied light emitted into the phonon-sideband (PSB) after excitation with resonant light pulses. For quantum information applications, however, it is important to extract the emitted zero-phonon line (ZPL) photons with high efficiency. In the current configuration, the excitation light is directly transmitted to the detector. Therefore, we separate out the ZPL photons from the bright excitation pulse with cross-polarization and time-bin filtering~\cite{Bernien2013,Hensen2015}, and shorten the excitation pulse further by introducing an additional electro-optic modulator (see supplementary material~\cite{supplement} for details). 

To be able to detect ZPL photons after a resonant excitation pulse, cross-polarization detection is especially important, because state of the art photodetectors have a dead time longer than the lifetime of the NV center; if a photon from the excitation pulse hits the detector, the dead time prevents detection of a ZPL photon, effectively reducing detection efficiency. Unfortunately, cross polarization only reduces the pulse power by a factor of 4 in our setup, which is likely due to vibrations of the freely hanging single mode fiber in the cryostat. We additionally insert a 8.6 dB attenuation neutral-density (ND) filter into the ZPL path so that the efficiency of the detector remains high. 

\figref[a]{ZPLandLosses} displays the fluorescence counts after a resonant excitation pulse, recorded simultaneously in the PSB and ZPL collection paths. The fluorescence in the ZPL path decays with a lifetime that agrees to within error with the lifetime of the NV center in the PSB path~\cite{supplement}. Thus, our technique enables us to see ZPL light from an NV center in a cavity under resonant excitation.

\begin{figure}[t!]
    \centering
    \includegraphics[width=0.47\textwidth]{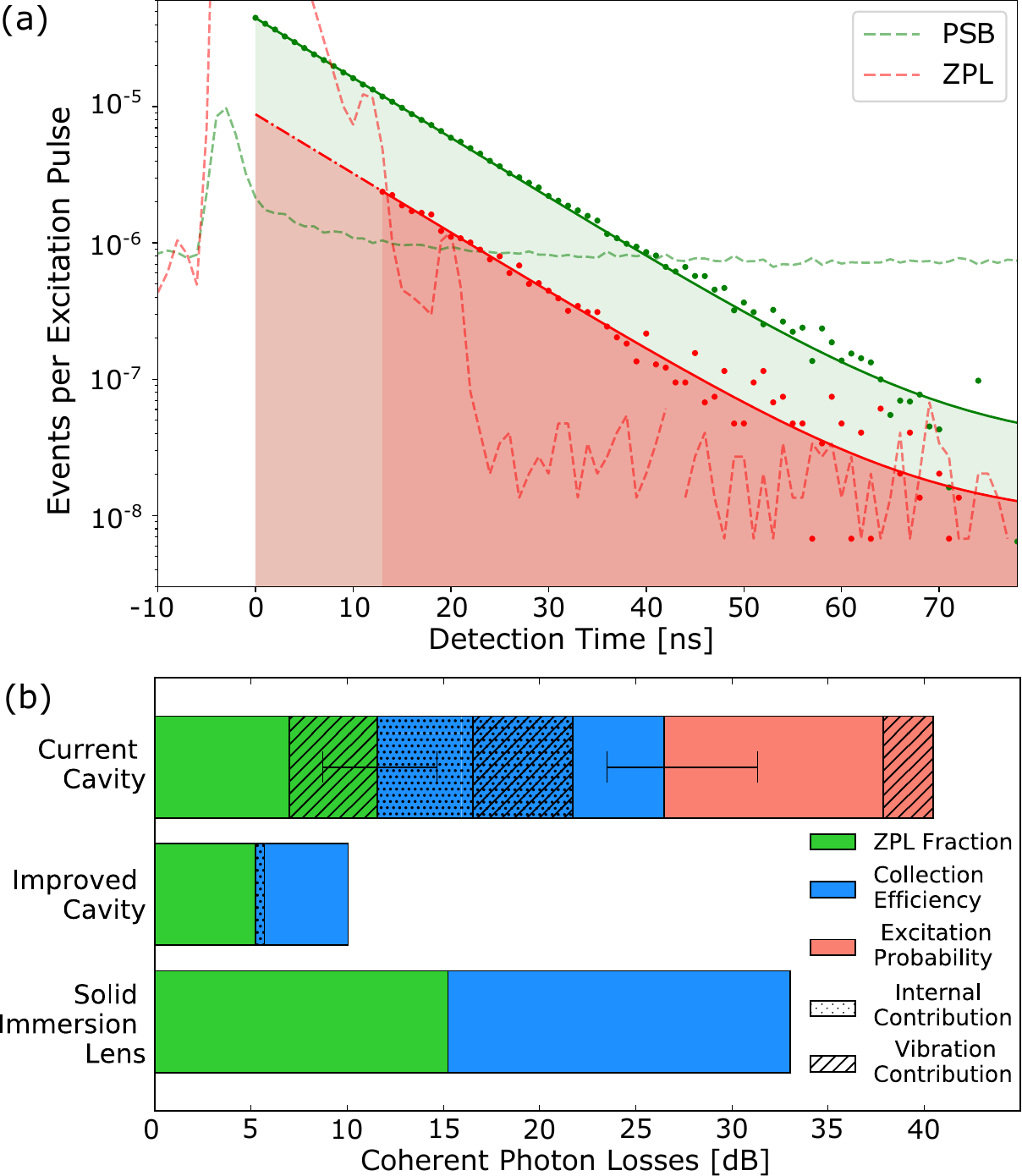}
    \caption{Zero-phonon line (ZPL) and phonon-sideband (PSB) fluorescence counts after resonant excitation and estimation of current and future ZPL count loss sources. (a) Background corrected PSB (green) and ZPL (red) fluorescence per excitation pulse (points), recorded at the same time. Background from the bright excitation pulse is shown with dashed lines~\cite{supplement}. The parallel decay of an exponential fit to both curves (solid lines) indicates emission from the same NV center. Shaded regions indicate areas for which fluorescence counts are summed to extract counts per excitation pulse (for the ZPL we include the extrapolated region shaded in light red)~\cite{supplement}. (b) Schematic of the sources of loss of coherent ZPL photons for the current cavity, the proposed improved cavity and a state-of-the art solid immersion lens (SIL) collection system. We breakdown the loss contributions into ZPL fraction, collection efficiency and excitation probability and separate the losses into vibration induced losses (stripes) and internal cavity losses (dotted) (see App.~\ref{sec:challenges}). }
    \label{fig:ZPLandLosses}
\end{figure}

Summing the fluorescence counts from the NV center in the ZPL path gives us a benchmark for the performance of our system as a collection enhancement tool. The excitation pulse obscures the initial counts, so we extrapolate a fit to the lifetime to extract the total NV center counts in the ZPL and correct for the ND filter. We find (9.3 $\pm$ 0.2) $\times10^{-5}$ photons per pulse in the ZPL path and (4.6 $\pm$ 0.1) $\times10^{-4}$ in the PSB path. Due to low available laser power, we operate with both low initialization probability into the m$_s$ = 0 state and low excitation probability to the excited state~\cite{supplement}.

To better compare the cavity with other collection enhancement systems, and to understand current limitations and possible improvements in a future system, we break the loss of coherent photons down into its main contributions, see \figref[b]{ZPLandLosses} (App.~\ref{sec:challenges} contains a detailed derivation and full discussion of these values). For the current cavity system, $\sim$ 14 dB of losses are associated with our limited probability of exciting the NV center to its excited state from one resonant laser pulse, $\sim$ 15 dB of losses result during photon collection, and $\sim$ 12 dB of losses result from the NV center emitting a photon in the PSB, rather than the ZPL. 

Importantly, correcting for the excitation probability alone already raises the ZPL detection probability to $\sim$ 2.0$\times10^{-3}$, which is comparable to the collection of coherent ZPL photons achieved for a NV center in a solid immersion lens ($\sim$ 5 $\times10^{-4}$)~\cite{Hensen2015,Humphreys2018}. Thus, despite the relatively low collection efficiency from our current cavity, limited in large parts due to operation in the clipping regime of the fiber mirror, our system already produces ZPL photons at a level comparable to state of the art non-cavity systems (see App.~\ref{sec:challenges}).

There are a number of changes which could improve ZPL collection under resonant excitation; we focus on three main developments which have already been achieved in other systems. First, introducing microwaves and a spin pumping laser is a standard technique for initializing and controlling the NV$^{-}$ charge and m$_s$ = 0 spin state with high fidelity in bulk samples. The narrow spectral diffusion linewidths demonstrated in this work should make this possible in our system as well. Together with upgraded laser excitation pulse power, this should allow for near unity excitation per pulse, as is common for bulk diamond samples~\cite{Robledo2010}. By increasing the polarization extinction by a factor of at least 100 by either fixing the fiber in the cryostat or switching to polarization maintaining fiber, the excitation pulse can be suppressed sufficiently even for these higher excitation powers. Second, reducing the vibrations by a factor of 20 (from $\sim$ 0.2 nm to $\sim$ 0.01 nm), as demonstrated in Refs.~\cite{Merkel2020,Casabone2020,Greuter2014,Brachmann2016,Janitz2017}, would increase the ZPL detection fraction by a factor of $\sim$ 16. Our current data already shows evidence for this potential improvement: by correlating the lifetime and the PSB fluorescence counts per resonant excitation pulse with the vibration level in the cryostat, we observe a reduction in lifetime (increase in fluorescence counts) from (10.02 $\pm$ 0.12) ns ([2.5 $\pm$ 0.1] $\times 10^{-4}$) to (9.77 $\pm$ 0.08) ns ([5.8 $\pm$ 0.2] $\times 10^{-4}$) for data collected during high and low vibration time periods of the cryostat, respectively. We explore this correlation in detail in App.~\ref{section:vib}. Finally, by working with a different fiber that is not clipping-loss limited, we expect to improve collection by a factor of 3. Together, these three improvements would raise the joint probability of producing and detecting a ZPL photon after short pulsed excitation to $\sim$ 10\%.

\section{CONCLUSION}

We have demonstrated the resonant operation of Purcell enhanced, coherent single photon emitters coupled to a fiber-based cavity system in a closed-cycle cryostat operated at a temperature of 4 K. We are able to address single Purcell enhanced NV centers via frequency-selective resonant excitation, and we have developed a sequence that allows us to collect up to (9.3 $\pm$ 0.2) $\times10^{-5}$ photons per excitation pulse in the zero-phonon line (ZPL). We have developed a theoretical model that describes our results, and used it to identify low excitation probability per laser pulse, length fluctuations between the cavity mirrors, and losses related to clipping on the fiber mirror as the main limitations in our current system. Using mutually non-exclusive numbers that have already each been achieved in several systems, we predict that we can increase the collected ZPL photons per excitation pulse to $\sim$ 10\% in a future NV-cavity system operated in a closed-cycle cryostat.

Building on the previous success of NV centers in entanglement generation and other network protocols~\cite{Bernien2013,Kalb2017,Humphreys2018}, a cavity enhanced NV-photon interface could dramatically improve entanglement rates and fidelity; single click and double click protocols would speed up by a factor of 100 and 10000 respectively. We expect that the realization of fully coherent quantum emitters embedded in optical fiber-based cavities will enable more extensive quantum networks with long coherence times, a crucial step towards a quantum internet.

\section{Acknowledgements}

We thank Wouter Westerveld, Martin Eschen, Guus Evers, and Santi Sager La Ganga for experimental assistance, Thomas Fink for fabrication of the fiber mirror, Lennart van den Hengel for electron irradiation of the diamond and Simon Baier and Conor Bradley for reviewing the manuscript. We acknowledge financial support from the EU Flagship on Quantum Technologies through the project Quantum Internet Alliance, from the Netherlands Organisation for Scientic Research (NWO) through a VICI grant, and the European Research Council (ERC) through an ERC Consolidator Grant.

\appendix

\section{CAVITY CHARACTERIZATION MEASUREMENTS} \label{sec:cav_char}

To characterize the cavity, we first input white light through the fiber, and measure the cavity transmission on a spectrometer for different cavity lengths. \figref[a]{finesse} displays the resulting dispersion diagram with avoided crossings between air and diamond modes~\cite{Janitz2015,Bogdanovic2017}. From a fit to this data, we infer that we operate in an air-like mode with a diamond thickness of 5.8 \um and a typical air gap between 5 \um and 7.5 \um (7.3 \um in the measurement shown)~\cite{Janitz2015}. This gap cannot be reduced further, likely due to an angled mounting of, or dirt on, the fiber. To measure the cavity linewidth, we then scan the cavity length through a resonance and measure the corresponding transmission peak after the cavity with a photodiode (sub-ms timescale to minimize vibration contributions to the linewidth). We apply sidebands to the laser to calibrate the frequency of this scan. In the measurement displayed in \figref[b]{finesse}, the linewidth of the cavity is $\kappa/2\pi$ = (3.5 $\pm$ 0.2) GHz, and the finesse is (2200 $\pm$ 100). Over the course of this work, we operated with a finesse between 1000 and 2500, dependent on the cavity position.

Based on the measured parameters of the cavity and a complete transfer matrix method described in Ref.~\cite{VanDam2018}, we can estimate the maximum possible Purcell enhancement in our cavity ($F_P^{ZPL} \approx$ 7), which corresponds to a coupling between NV centers and the cavity of $g/2\pi\approx$ 300 MHz. This estimate holds for an NV center at the optimal depth location, which is a cavity standing wave antinode, and the optimal xy position for maximum overlap with the Gaussian mode of the cavity. We describe the mismatch with the ideal position with a parameter, $\xi$, which we measure via the Purcell enhancement in the main text. The coupling is also reduced by vibrations as discussed in App.~\ref{section:theory}. The lifetime limited transition linewidth of the NV center is $\gamma/2\pi\approx$ 13 MHz~\cite{Robledo2011}. In Sec.~\ref{section:Off-resonant} and Sec.~\ref{sec:Resonant_purcell}, we measured the Purcell enhancement under off-resonant and resonant excitation and determined $F_P^{ZPL}\approx$ 4 and thus $g/2\pi\approx$ 180 MHz (see eq.~\ref{eq. cooperativity}). The relative values of $g,\kappa$ and $\gamma$, put us firmly in the weak coupling Purcell regime ($\gamma\ll g\ll\kappa$), which is the ideal parameter range for collecting photons from an emitter~\cite{Reiserer2015}.

\begin{figure}
    \raggedright
    \includegraphics[width=0.47\textwidth]{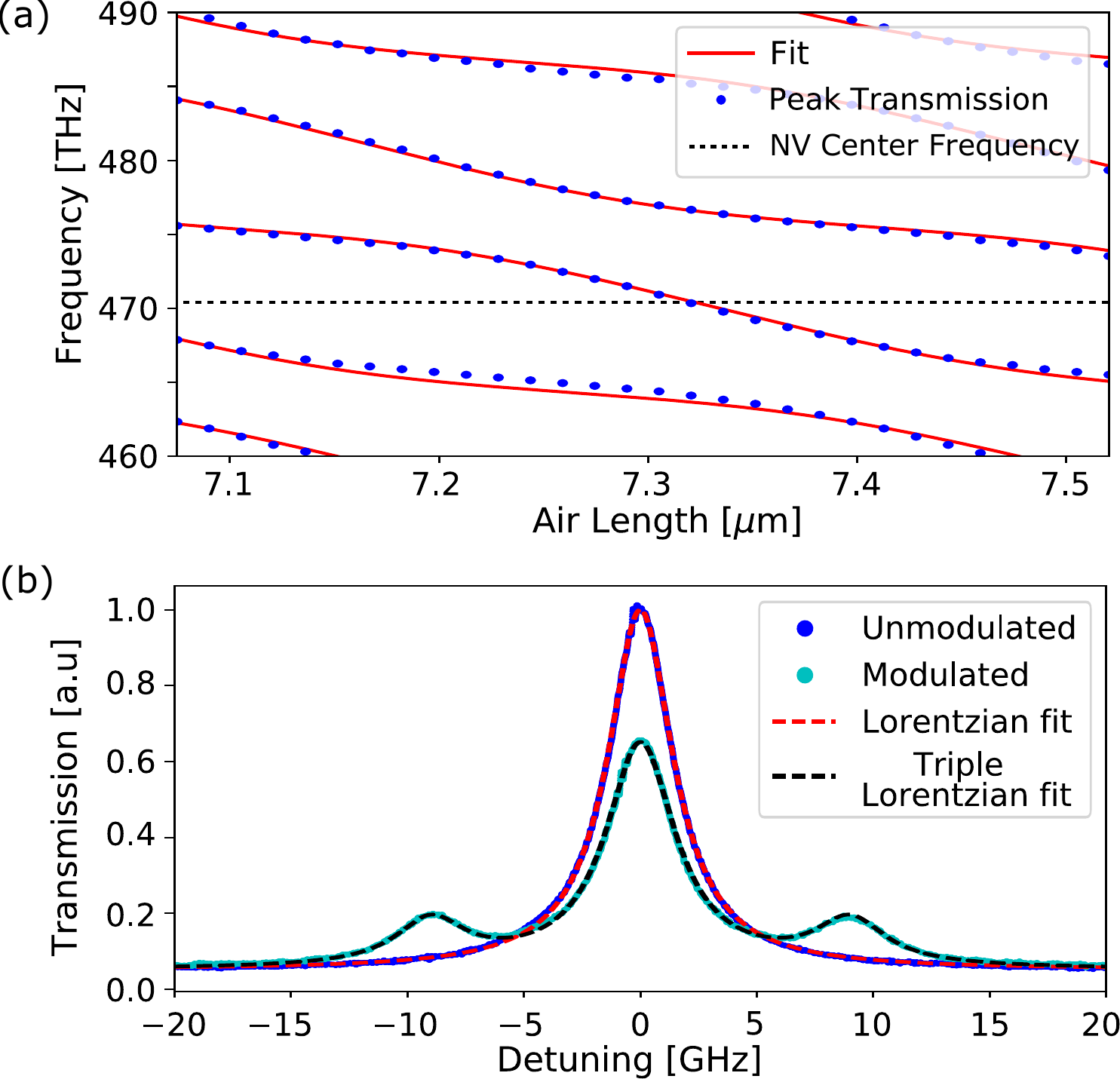}
    \caption{Example data traces of cavity dispersion and finesse. (a) Cavity dispersion for the coupled membrane-air fiber-cavity system. The length of the cavity is swept and transmission is recorded on a spectrometer. From fits of the fundamental modes to a transfer matrix model, we can determine the diamond and air thicknesses to be 5.8 \um and 7.3 \umdot, respectively. (b) Finesse measurement. We scan the cavity quickly over a transmission peak to extract the cavity linewidth ($\kappa/2\pi$ = [3.5 $\pm$ 0.2] GHz) from fits to the data; the laser exciting the cavity is modulated with sidebands of 9 GHz with an EOM, serving as frequency calibration reference. }
    \label{fig:finesse}
\end{figure}

\section{THEORETICAL MODEL} \label{section:theory}

\subsection{Purcell Enhancement}

NV centers exhibit limited emission into the zero-phonon line (ZPL), $\beta_0$, given by the Debye-Waller factor. Some states also couple to a non-radiative transition through an intersystem crossing~\cite{Doherty2013}. Without interaction with a cavity, we can model the decay from the NV center excited state, $\gamma_0$, with a radiative decay rate, $\gamma_{rad}$, and non-radiative decay rate, $\gamma_{dark}$, as:
\begin{equation}
\gamma_0 = \beta_0\gamma_{rad} + (1-\beta_0)\gamma_{rad} + \gamma_{dark}.
\label{eq. nonsimplifiedcavdecay}
\end{equation}
Coupling to the mode of the cavity opens up another decay channel for the NV center, which we characterize with a Purcell factor $F_P^{ZPL}$. The decay rate from the NV excited state, modified by the cavity, $\gamma'$, is then given as
\begin{equation}
\gamma' =  F_P^{ZPL}\beta_0\gamma_{rad}+ (1-\beta_0)\gamma_{rad} + \gamma_{dark}, 
\end{equation}
with
\begin{equation}
F_P^{ZPL} = \frac{4g^2}{\kappa\gamma} + 1, 
\label{eq. cooperativity}
\end{equation}
where $g/2\pi$ is the NV-cavity coupling, $\kappa/2\pi$ is the cavity linewidth, and $\gamma/2\pi$ is the lifetime limited optical transition linewidth of the NV center~\cite{Riedel2017a}. We choose the definition of the Purcell factor as the enhancement of the ZPL --- rather than the enhancement of the total emission from the NV center --- because it better reflects the increase in coherent light.

In this work, we primarily investigate transitions with linear polarization, and we preferentially initialize into the $m_s$ = 0 ground state with green excitation. The transitions in the $m_s$ = 0 manifold are $E_x$ and $E_y$, for which $\gamma_{dark}$ is much smaller than $\gamma_{rad}$ in bulk diamond ~\cite{Goldman2015a, Kalb2018}: In Sec.~\ref{sec:Resonant_purcell}, we measure a reduced lifetime when the cavity is off resonance of (10.9 $\pm$ 0.2) ns, compared to the $\sim$12 ns typically found in bulk diamond ~\cite{Goldman2015a, Kalb2018}; from this, we estimate that $\gamma_{dark}/\gamma_{rad}\approx0.1$, so we assume $\gamma_{rad}\gg\gamma_{dark}$, and simplify eq.~\ref{eq. nonsimplifiedcavdecay} to 
\begin{gather}
\frac{\gamma'}{\gamma_0} = \frac{\tau_0}{\tau'} = 1 + \left(F_P^{ZPL}-1\right)\beta_0\frac{\gamma_{rad}}{\gamma_{rad}+\gamma_{dark}} \\ \approx 1 + \left(F_P^{ZPL}-1\right)\beta_0, \nonumber
\end{gather}
where $\tau_0$ is the NV center lifetime without the cavity and $\tau'$ is the modified lifetime. The fraction emitted into the ZPL in the cavity mode is $(F_P^{ZPL}-1)\beta_0/(F_P^{ZPL}\beta_0+1-\beta_0)\approx(F_P^{ZPL}-1)\beta_0$. If the assumption that $\gamma_{dark}$ is negligible is incorrect, then $(F_P^{ZPL}-1)\beta_0$ is increased by a factor of $(\gamma_{rad}+\gamma_{dark})/\gamma_{rad}$. A simple rearrangement then gives us eq.~\ref{eq:FZPL}.

The rates of detected photons in the ZPL path, $C_{ZPL}$, and the phonon-sideband (PSB) path, $C_{PSB}$, are
\begin{gather}
\label{eq:ZPL_general}
C_{zpl}(t) = p_{ex} \frac{\left(F_P^{ZPL}-1\right)\beta_0}{F_P^{ZPL}\beta_0+(1-\beta_0)} \eta_{zpl} \gamma'e^{-\gamma't} \\
= p_{ex}\left(F_P^{ZPL}-1\right)\beta_0 \eta_{zpl} \gamma_0e^{-\gamma't} \nonumber
\end{gather}
and
\begin{gather}
C_{psb}(t) = p_{ex} \frac{1-\beta_0}{F_P^{ZPL}\beta_0+(1-\beta_0)} \eta_{psb} \gamma'e^{-\gamma't} \\
= p_{ex} (1-\beta_0) \eta_{psb} \gamma_0e^{-\gamma't} \nonumber,
\label{eq:gen_counts}
\end{gather}
where $p_{ex}$ is the probability that a pulse excites the NV center to its excited state, and $\eta_{zpl}$ ($\eta_{psb}$) is the detection efficiency for ZPL (PSB) photons from the cavity. A number of these parameters depend on the cavity transmission, $T$, which itself depends on the detuning between the cavity and the transition, $\Delta$:
\begin{equation}
T(\Delta) = \frac{\kappa^2/4}{\kappa^2/4 + \Delta^2}.
\end{equation}
This modifies $F_P^{ZPL}$ and $\eta_{zpl}$ to
\begin{gather}
F_P^{ZPL}(\Delta) = \left(F_P^{ZPL}(0)-1\right)T(\Delta)+1 \\
\eta_{zpl}(\Delta) = \eta_{zpl}(0)T(\Delta)
\end{gather}

The PSB collection is not resonant with the cavity, and its spectral distribution lies mostly outside of the stopband of the mirror, so there should be minimal changes from small variations in cavity detuning. For the same reason, green excitation light is also independent of the cavity detuning. However, resonant red excitation depends strongly on cavity detuning, and the corresponding power in the cavity. The Rabi frequency is proportional to the square root of the intracavity power which scales with $T(\Delta)$. Therefore, in the resonant excitation case:
\begin{equation}
p_{ex}(\Delta) = \sin(\phi_{p}(0)\sqrt{T(\Delta)}/2) 
\end{equation}
Here $\phi_{p}$(0) is the Rabi rotation angle induced by the pulse on resonance (e.g.~$\pi$ for a complete population of the excited state). In the weak excitation limit explored in this work, this can be approximated as $\phi_{p}(0)\sqrt{T(\Delta)}/2$.

\subsection{Vibration Model}
The above simple model does not take into account the vibrations in the cavity length. We extend the model of Ref.~\cite{VanDam2018} to build a complete numerical transfer matrix model for the cavity. We assume a Gaussian distribution of cavity lengths, $f_{vib}$, given as
\begin{equation}
f_{vib}(dL) = \frac{1}{\sqrt{2\pi\sigma^2}}e^{-\frac{dL^2}{2\sigma^2}} \label{eq:len_dist},
\end{equation}
where $\sigma$ is the width of the length distribution, and $dL$ the length flucutations around a certain cavity length, as induced from vibrations. We then find the Purcell-enhanced fraction of emission into the ZPL ($F_p(dL)\beta_0$) for each dL. 

\subsubsection{Off-resonant Excitation}

Under off-resonant excitation, the probability of exciting the NV center does not depend on the length of the cavity. For each frequency detuning, $\Delta$, we find the equivalent cavity detuning length, $L_{det}$, using our complete model of the cavity dispersion relationship. We integrate eq.~\ref{eq:ZPL_general} over the length distribution in eq.~\ref{eq:len_dist} to determine the emitted counts in the ZPL:
\begin{widetext}
\begin{gather}
C_{zpl}(t,L_{det}) = p_{ex}\gamma_0\int dL \text{ } f_{vib}(dL)g_{nv}(L_{det}+dL)\beta_0\left[F_P^{ZPL}(L_{det}+dL)-1\right]\eta_{zpl}(L_{det}+dL)\exp\left[-\gamma'(L_{det}+dL)t\right]  \\
 = p_{ex}\beta_0\left[F_P^{ZPL}(0)-1\right]\eta_{zpl}(0)\gamma_0\int dL \text{ } f_{vib}(dL)g_{nv}(L_{det}+dL)T(L_{det}+dL)^2\exp\left[-(1-\beta_0+\beta_0F_P^{ZPL}(L_{det} + dL))\gamma_0 t\right]. \nonumber 
 \label{eq.off_res_zpl}
\end{gather} 
\end{widetext}
We then numerically integrate and fit the function $C(t)$ to an exponential decay to determine the measured lifetime $\tau(\Delta)$. We also sum over $t$ to find total counts $C_{tot}(\Delta)$. For the case of multiple NV centers we also introduce an extra Gaussian broadening term, $g_{nv}$, inside the integral. The exact form of $g_{nv}$ depends on the (unknown) distribution of NV centers, but we find that the Purcell factor we extract is relatively insensitive to the shape we pick.

\subsubsection{Resonant Excitation} \label{sec.:res_exc_theo}

Under resonant excitation the probability of exciting the NV center depends strongly on the excitation laser detuning from the cavity. This modifies the function for counts from the cavity slightly. We integrate the counts curves in the same way as in the case of off-resonant excitation to determine the total counts. For collection from the ZPL and the PSB, respectively, we get
\begin{widetext}
\begin{gather}
    C_{zpl}(t,L_{det})=\frac{p_{in}\phi_p(0)}{2}\beta_0\left[F_P^{ZPL}(0)-1\right]\eta_{zpl}(0)\gamma_0\int dL \text{ } f_{vib}(dL)T(L_{det}+dL)^{5/2}\exp\left[-(1-\beta_0+\beta_0F_P^{ZPL}(L_{det} + dL))\gamma_0 t\right] \label{eq:ZPL_resonant_counts},\\
    C_{psb}(t,L_{det}) = \frac{p_{in}\phi_p(0)}{2}(1-\beta_0)\eta_{psb}\gamma_0\int dL \text{ } f_{vib}(dL)\sqrt{T(L_{det}+dL)} \exp\left[-(1-\beta_0+\beta_0F_P^{ZPL}(L_{det} + dL))\gamma_0 t\right].
    \label{eq.res_psb}
\end{gather}
\end{widetext}
$p_{in}$ is the probability of initializing in the $m_s$=0 state. We do not include $g_{nv}$ because we only excite single NV centers in the case of resonant excitation. Counts and lifetime are then calculated in the same manner as for the off-resonant case. 

\section{VIBRATIONS AND THEIR INFLUENCE ON COUNTS AND LIFETIMES} \label{section:vib}

All measurements in this paper were taken in a closed-cycle cryostat operated at $T \sim 4$ K. Such a cryogenic systems has intrinsic vibrations due to moving parts. Nevertheless, we chose a closed-cycle system - as opposed to essentially vibrations free liquid helium bath cryostats - because of its ease of operation, and the possibility of uninterrupted measurement cycles on a timescale of several months without any human intervention; this is an important operational consideration for a future quantum network, with nodes distributed over distant locations, as it removes the cost and labour associated with a helium infrastructure.

The average root mean square (rms) fluctuations in cavity length during one cryostat period, inferred from a cavity transmission measurement (see supplemental materials~\cite{supplement} for measurement details and additional data), are displayed in \figref[a]{lt_dep_vib}. We observe a (5 - 6) fold change in vibration level over the course of a coldhead cycle, with two spikes in the data, related to the coldhead starting to move up and down, respectively. Therefore, the vibrations in the system are directly linked to the coldhead movement.

As discussed in App.~\ref{section:theory}, cavity length fluctuations influence the collected counts and detected lifetime from an NV center. In the case of resonant excitation and PSB detection, for example, both the excitation probability, and the Purcell enhancement, depend on the cavity detuning at a given time. To probe this dependence, we utilize a synchronization signal from the cryostat coldhead that marks the beginning of a new period, and use it to record the time in the cryostat period that a NV center photon is detected. We assign NV center photon arrival timestamps to periods in the cryostat with low (orange shaded area in \figref[a]{lt_dep_vib}) and high vibration values (blue shaded area in \figref[a]{lt_dep_vib}). We can then extract the vibration influence on counts and corresponding excited state lifetime, see \figref[b]{lt_dep_vib}. During the cryostat times with low vibrations, we extract a NV center lifetime (counts per excitation pulse) of (9.77 $\pm$ 0.08) ns ([5.8 $\pm$ 0.2] $\times 10^{-4}$), compared to (10.02 $\pm$ 0.12) ns ([2.5 $\pm$ 0.1] $\times 10^{-4}$) during high vibration times. In comparison, the lifetime (counts) over all cryostat times is (9.87 $\pm$ 0.04) ns ([4.7 $\pm$ 0.2] $\times 10^{-4}$). This serves as a further proof that it is indeed Purcell enhancement through coupling to a cavity that leads to a lifetime reduction and increase in counts of the NV center when the cavity frequency is tuned to the NV center transition frequency. Additionally, it also shows that reducing the vibrations in a future system increases Purcell enhancement, as discussed in App.~\ref{sec:challenges} below. Note that we average over all cryostat period times in the measurements in this study, except if stated otherwise.

\begin{figure}
    \raggedright
    \includegraphics[width=.485\textwidth]{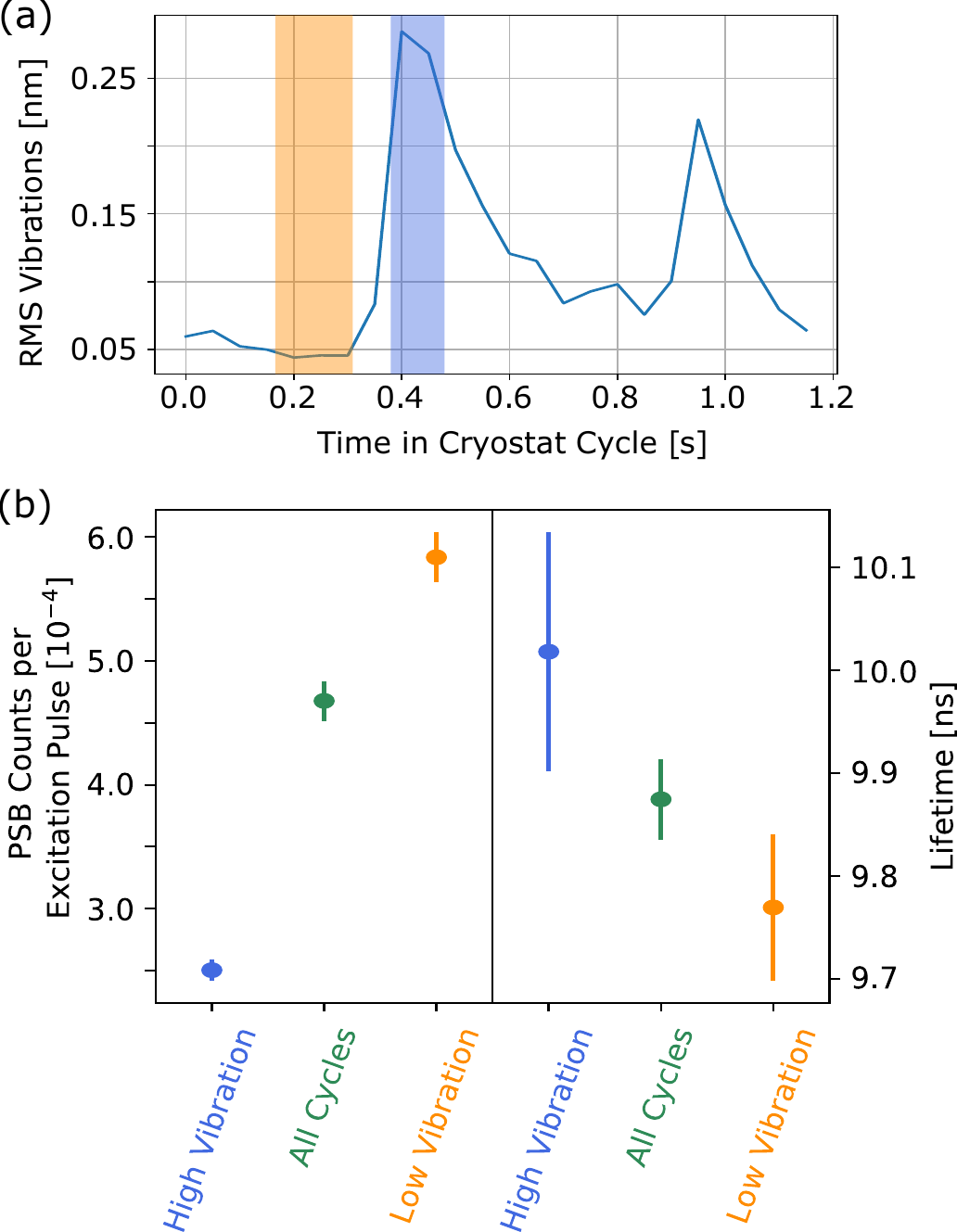}
    \caption{Cavity vibrations cycle and its influence on collected NV center counts and lifetime. (a) Using a synchronization signal of the cryostat marking the beginning of a new coldhead cycle, the root mean squared (rms) vibrations of the cavity, inferred from a cavity transmisison measurement, are averaged with respect to time in the cryostat cycle. For measurement and analysis details, see the supplemental materials~\cite{supplement}. Blue and orange shaded regions indicate time periods assigned as having low and high vibration values for data analysis similar to the one in (b). (b) Influence of the vibrations cycle on PSB counts per excitation pulse (left side) and lifetime (right side) for the case of resonant NV center excitation. The cavity is kept on resonance with the NV center transition frequency. Note that this data was taken for a different cooldown and thus vibration level assignment as (a).}
    \label{fig:lt_dep_vib}
\end{figure}

\section{CURRENT CHALLENGES AND FUTURE IMPROVEMENTS} \label{sec:challenges}

In Sec.~\ref{sec:ZPL_collection}, we measured the absolute counts from an NV center in the cavity in both the PSB and the ZPL under resonant excitation. It is helpful to understand the loss contributions which prevent perfect PSB and ZPL detection. In this way, we can project what improvements may be possible with future upgrades to the setup. The investigations in this section underlie the loss plot in~\figref[b]{ZPLandLosses}.

\subsection{Vibrations}

First, we estimate the effect of vibrations on cavity coupling and enhancement. In App.~\ref{section:vib} and the supplementary materials~\cite{supplement}, we characterize the vibrations in the system. In our model in App.~\ref{section:theory}, the vibrations reduce three parameters: the excitation probability of the NV center, the ZPL fraction emitted into the cavity, and the collection efficiency of the cavity. These effects are correlated, because the vibrations are slow ($\sim$ ms timescale) compared to the excitation and emission timescales of the NV center ($\sim$ ns timescale). Only the excitation probability of the NV center has a notable contribution to the PSB counts, as the fraction into the PSB is only slightly decreased for our cavity, and the PSB is not collected resonantly with the cavity.

From the data in~\figref[b]{ResonantPurcell} and~\figref[a]{ZPLandLosses} of the main text, we determine that (7.0 $\pm$ 3.4)\% of the emission goes into the ZPL path, and we excite and detect ZPL photons with a probability of (9.3 $\pm$ 0.2) $\times10^{-5}$. If we set the vibration level to zero in our model, we find that 17$\%$ is emitted into the ZPL, and 1.7x$10^{-3}$ ZPL photons are collected. Thus, the total reduction of ZPL collection and excitation due to vibrations is 13 dB. Because these reductions are correlated, we distribute the vibration contributions to the losses in~\figref[b]{ZPLandLosses} according to the relative contribution to the $T^{5/2}$ term in eq.~\ref{eq:ZPL_resonant_counts}.

\subsection{Collection Efficiency}

We divide the ZPL collection efficiency into two parts: internal collection efficiency and external collection efficiency. The external collection efficiency is the coupling between the mode exiting the cavity and the detector, and is determined by the classical optics in between. Internal collection efficiency is the probability that a photon escapes the cavity and couples into the free space mode. 

The internal efficiency, $\eta_{int}$, can be estimated using the measured transmission rate of the free space mirror, $\kappa_{FS}$, as $\eta_{int} = \kappa_{FS}/\kappa$. We design $\kappa_{FS}$ to be significantly larger than the fiber transmission and scattering loss rates. However, the angle of the fiber forced us to operate in a regime where diffraction losses contributed significantly to the cavity finesse (which can be readily overcome with a new fiber). Therefore, in this work we operated with $\eta_{int}$ in the range of 0.05 - 0.17, which also includes a factor of $\sim 1/3$ from vibrations. 
\begin{figure}
    \raggedright
    \includegraphics[width=.475\textwidth]{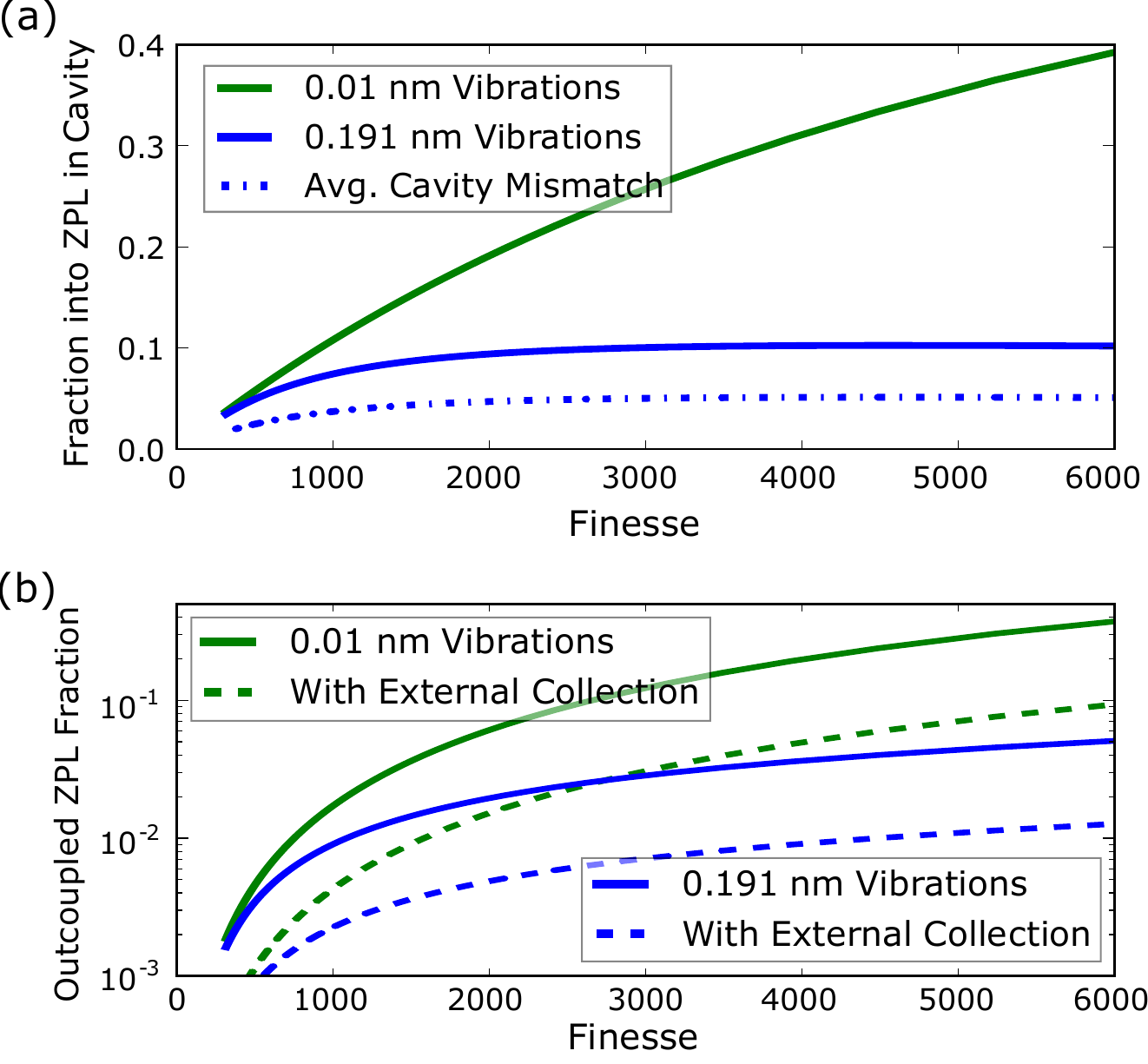}
    \caption{Projections for ZPL emission and collection fraction. We fix the design parameters of the mirrors to match our current system (Finesse 6200). Fraction emitted into the ZPL (a) and outcoupled fraction of photons in the ZPL (b) for the current (blue) and improved (green) vibration levels, as a function of achieved cavity finesse. These simulations are based on the model from Ref.~\cite{VanDam2018}.}
    \label{fig:projections}
\end{figure}

To measure the external collection efficiency, we send classical light through the fiber, and estimate the losses at each section between the fiber input and the detector. The loss to and from the cavity, probed in the fiber in reflection, is approximately -12.4 dB, including the fiber connector and fiber splice twice, and the reflection off the cavity once. Based on the depth of the reflection dip from the cavity, we estimate that we have a 2.8\% incoupling efficiency. Based on the design values of our mirrors, we expect 2.5\% incoupling from the fiber side, which is in reasonable agreement, and suggests that there is not a large mode mismatch between the cavity and the fiber. For a measurement of the transmitted light, we filter the input laser beam with a 20 dB filter and measure counts on an APD in the ZPL path. Subtracting out the light lost in the excitation path, we estimate that the total collection efficiency is 4\%. This is likely divided into $\sim$ 11\% internal collection efficiency and $\sim$ 32\% external collection efficiency, but the precise contributions are not known. The external efficiency is determined by classical optics and the detector efficiency, so we expect that the external efficiency can also be improved in future experiments with better coatings and superconducting nanowire detectors.

We can also estimate the collection efficiency of the PSB path to make full use of the direct comparison in counts between PSB and ZPL. This efficiency is relatively low in our setup, because we have a long working distance objective with a numerical aperture of 0.55 (the supplementary material~\cite{supplement} contains a full description of the optics path). Depending on the angle of the NV center dipole emission axis in our $<$100$>$ sample, the collection is between 0.9\% and 2.5\%. There is likely a bias towards NV centers with a better coupled dipole, because they produce more collectable PSB counts, and thus are easier to spot when searching for NV centers. The top collecting mirror has a narrow stop band, which by design allows 83\% of the PSB light through. We estimate 83\% transmission through the path and 70\% APD detection efficiency, but the exact values are unknown. Altogether we expect between 0.4\% and 1.2\% collection efficiency in the PSB.

\subsection{Benchmarking with Excitation Probability Correction}

\begin{table}[t!]
    \centering
    \begin{tabular}{p{3.5cm}|l|p{1.4cm}|p{1.4cm}}
        Improvement & Achieved in & Enhance -ment Factor & ZPL Collection (\%) \\ \hline\hline
        Spin Pump / Resonant Repump & \cite{Robledo2011,Bernien2013} & 20 & 0.2\\ & & & \\
        20x Vibration Reduction & \cite{Merkel2020,Casabone2020,Greuter2014,Brachmann2016} & 16 & 3\\ & & &\\
        Finesse limited by mirror coatings (6000) & \cite{Bogdanovic2017,Jensen2020} & 3 & 9\\ \hline
        Higher Finesse (11000) & \cite{Bogdanovic2017,Jensen2020} & 1.2 & 10 \\  & & &\\
        Diamond-like Mode & \cite{Jensen2020,Salz2020,Bogdanovic2017} & 2 & 20
    \end{tabular}
    \caption{Summary of suggested improvements to current diamond fiber-cavity. The first three upgrades represent improvements that have already been achieved in other systems, as indicated in the second column. Finesse would be limited by the mirror if achieved losses limit the finesse to significantly greater than 6000. The last two items have been achieved but not simultaneously with the other requirements. We give the resulting enhancement of the ZPL collected and the absolute fraction collected in the ZPL.}
    \label{tab:future}
\end{table}

We can benchmark our system against other diamond collection optics such as solid immersion lenses. However, the state of the art in these systems is near perfect initialization and excitation of the NV center. Higher initialization can be achieved with an additional spin pumping laser and microwaves~\cite{Bernien2013}. In our system we initialize with a green laser, which only has a limited probability of initializing into the $m_s$ = 0 spin state. Furthermore, we do not saturate the initialization with green or the excitation with red, due to limited laser power, so it is hard to estimate the probability of excitation. Therefore, the counts we observe are not directly comparable to state of the art.

We can instead use our calculations of the collection efficiencies to estimate the excitation probability based on the total counts we observe from the NV center in the PSB and the ZPL. First we use the ZPL counts; we expect that with perfect initialization the counts in the ZPL should be $(F_P^{ZPL}-1)\beta_0/(F_P^{ZPL}\beta_0+1-\beta_0)\times\eta_{int}\times\eta_{ext}$ or approximately 2.6x$10^{-3}$ based on the results of Sec.~\ref{sec:Resonant_purcell} and Sec.~\ref{sec:ZPL_collection}. This corresponds to an excitation probability of 3\%. For comparison, the PSB counts should be $(1-\beta_0)/(F_P^{ZPL}\beta_0+1-\beta_0)\eta_{psb}$. The estimated collection efficiency and measured counts give us an excitation probability between and 14\% and 4\% (with a bias towards the latter). This corresponds to between 0.6 and 2x$10^{-3}$ counts in the ZPL with perfect initialization. The estimates show reasonable agreement, and exact determination of the different losses is left for future work. Even with the limited collection efficiency in this work, the ZPL counts we collect are comparable to state of the art when correcting for initialization and excitation probability.

\subsection{Improvements}

Based on the current challenges and our modelling, we suggest three improvements which should be possible in the short term, as summarized in Tab.~\ref{tab:future}. We constrain ourselves to performances that have already been achieved in other systems. The first improvement is to implement microwave control of the ground state NV center spin, which will allow spin pumping and also yellow resonant repumping for higher initialization and even lower spectral diffusion. Using these techniques, it should be possible to achieve near unity excitation as has been shown in bulk diamond ~\cite{Robledo2011,Bernien2013}.

The second improvement is to reduce vibrations; a number of research groups have achieved considerably lower vibrations of comparable fiber-based micro-cavities ~\cite{Merkel2020,Casabone2020,Greuter2014,Brachmann2016}. In Fig.~\ref{fig:projections}, we simulate the dependence of ZPL emission fraction and collection fraction on the finesse for different vibration levels. We project that a factor of 20 vibration levels reduction gains a factor of 16 in total ZPL collection. 

The third improvement is increasing the finesse; our current finesse is limited by diffraction losses, but with a better fiber it should be limited by the mirror coatings, gaining another factor of 3. Although it has not yet been achieved simultaneously, a diamond fiber-cavity with finesse 11000 diamond-like modes would further increase ZPL collection by approximately a factor of 2. Altogether we expect that about 10\% ZPL collection should be possible with already achieved parameters and 20\% with combining the best results from different setups.

\bibliography{cav1}

\end{document}